# Boosting the superconducting spin valve effect in a metallic superconductor/ferromagnet heterostructure

Pavel V. Leksin[1,2] (✉), Andrey A. Kamashev[2], Joachim Schumann[1], Vladislav E. Kataev[1], Jürgen Thomas[1], Bernd Büchner[1,3], and Ilgiz A. Garifullin[2]

[1] *Leibniz Institute for Solid State and Materials Research Dresden (IFW Dresden), D-01171 Dresden, Germany*
[2] *Zavoisky Physical-Technical Institute, Russian Academy of Sciences, 420029 Kazan, Russia*
[3] *Technical University Dresden, D-01062 Dresden, Germany*

**ABSTRACT**

Superconducting spin valves based on the superconductor/ferromagnet (S/F) proximity effect are considered to be a key element in the emerging field of superconducting spintronics. Here, we demonstrate the crucial role of the morphology of the superconducting layer in the operation of a multilayer S/F1/F2 spin valve. We study two types of superconducting spin valve heterostructures, with rough and with smooth superconducting layers, using transmission electron microscopy in combination with transport and magnetic characterization. We find that the quality of the S/F interface is not critical for the S/F proximity effect, as regards the suppression of the critical temperature of the S layer. However, it appears to be of paramount importance in the performance of the S/F1/F2 spin valve. As the morphology of the S layer changes from the form of overlapping islands to a smooth case, the magnitude of the conventional superconducting spin valve effect significantly increases. We attribute this dramatic effect to a homogenization of the Green function of the superconducting condensate over the S/F interface in the S/F1/F2 valve with a smooth surface of the S layer.

It is well known that bringing together two antagonistic ordered states of matter, namely superconductivity (S) and ferromagnetism (F), in a thin S/F heterostructure gives rise to a variety of new physical phenomena, such as the S/F/S $\pi$-phase Josephson effect, the so-called cryptoferromagnetic state, conventional (singlet) and unconventional (triplet) superconducting spin valve effects (SSVE) (e.g., see a recent review [1] and references therein). The SSVE for a sequence of two metallic F layers and one S layer, S/F1/F2, was proposed theoretically in 1997 by Oh et al. [2]. This multilayer structure can be switched between the normal and superconducting states by changing the mutual orientations of the magnetizations of the F1 and F2 layers between parallel (P) and antiparallel (AP) configurations. The physical mechanism behind this effect involves manipulating the phase and amplitude of the superconducting wave function penetrating into the F1 layer and, hence, the superconducting critical temperature, $T_c$, by changing the magnetic state of the F1/F2 part of the heterostructure. A similar theory for a F1/S/F2 multilayer was proposed in 1999 by Tagirov [3] and Buzdin et al. [4]. Later, a triplet spin valve effect was described theoretically for S/F1/F2 structures by Fominov et al. [5–7], who proposed another way to manipulate $T_c$, which is related to the formation of a long-range triplet component of the superconducting condensate at a non-collinear orientation of the F1 and F2 magnetizations.

At present, there have been a number of experimental works confirming the SSVE (see, e.g.,

Address correspondence to Pavel V. Leksin, p.leksin@ifw-dresden.de

Refs. [8–10]). In most of these cases, the magnitude of the effect, $\Delta T_c = T_c^{AP} - T_c^{P}$, was of the order of 10–40 mK, whereas the width of the superconducting transition was $\delta T_c \approx 100$ mK (see references in Ref. [1]). Therefore, no full switching between the normal and superconducting states could be achieved. Finally, for the case of the S/F1/F2 multilayer, full switching due to the SSVE was realized by means of a notable reduction of $\delta T_c$ [11]. However, another (complementary) way to achieve full switching would be to increase $\Delta T_c$, since theories predict substantially larger magnitudes of the SSVE as compared to the values found experimentally thus far. One should note that, similar to an early work by Deutscher and Meunier [12], there have been several recent observations of a high $\Delta T_c$ value for F1/S/F2 and S/F1/F2 structures with insulating and half-metallic F-layers [13, 14]. However, the physical origin of such a significant effect has not yet been understood theoretically and its relation to the SSVE remains unclear.

Up to now, the role of the microscopic structure of the superconducting layer in S/F and S/F1/F2 proximity effects has received little attention from both theoretical and experimental researchers. Recent theories on S/F1/F2 proximity [5–7] treat the S layer as a flat and continuous layer without any more detailed discussion. Thus, in this work, we demonstrate experimentally that a significant cause of the small magnitude of the SSVE in metallic S/F1/F2 heterostructures is the rough surface of an S layer composed of overlapping islands, which can reduce $\Delta T_c$ to zero. By improving the morphology of the S layer to a smooth surface, we are able to significantly enhance $\Delta T_c$, up to 100 mK. This highlights the key role that the quality of the S layer in metallic heterostructures has in the SSVE related to S/F proximity.

In order to investigate the influence of the type of S layer structure on the S/F proximity effect, we prepared the following groups of samples: bilayer S/F heterostructures (Fig. 1(a)) and S/F1/F2-based spin valve samples (Fig. 1(c)). We also prepared a trilayer sample, S/AD/F (Fig. 1(b)) to demonstrate the importance of the antidiffusion (AD) layer, introduced between the S and F layers, in improving the quality of superconducting transitions without influencing the S/F proximity effect. Each of these groups had two types of S layer: (i) an S layer composed of overlapping islands, which will be hereafter called a rough S layer, and (ii) a smooth S layer (Fig. 1).

For the implementation of the S/F1/F2-based spin valve, we prepared samples with the layer sequence AF/F1/N/F2/AD/S deposited on a MgO(100) substrate (Fig. 1(c)). Here, the nonmagnetic metallic layer (N) between the F1 and F2 layers decouples the magnetizations of the F layers. The antiferromagnetic (AF) layer pins the magnetization of the F1 layer, whereas the magnetization of the F2 layer remains free. The materials chosen were as follows: for the F layers, we used permalloy, $Py = Ni_{0.81}Fe_{0.19}$; the N and

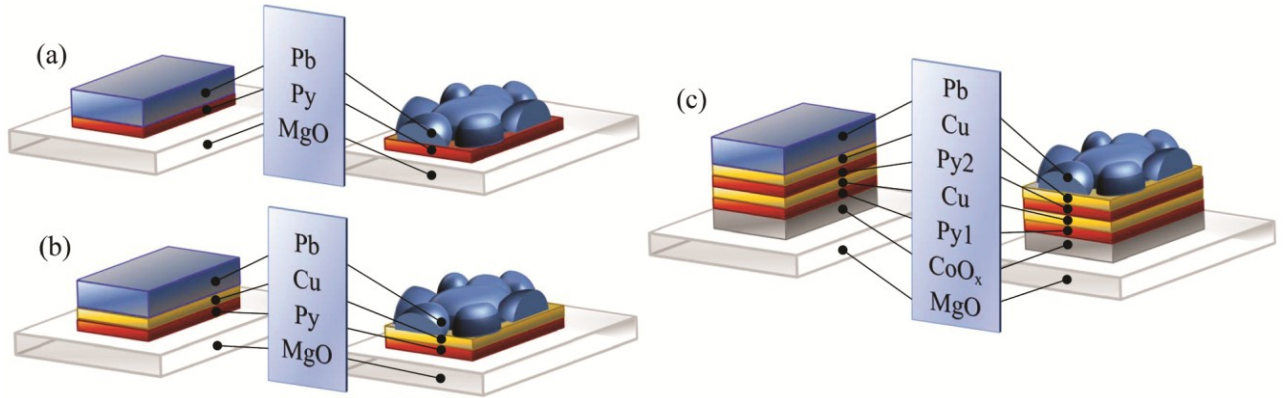

**Figure 1** Schematic design of the samples: (a) bilayer, (b) trilayer, and (c) spin valve structures with (right) rough and (left) smooth surfaces of the S film

AD layers were made of Cu; Pb was used for the S layer; and CoO$_x$ was used for the AF layer.

The deposition of layers was performed using an e-gun in ultra-high vacuum (UHV) with a pressure of $10^{-9}$ mbar. The deposition setup included a load lock station with vacuum shutters, allowing us to change the sample holder without breaking the UHV in the main deposition chamber. First, the substrates were fixed on a sample holder and transferred into the main deposition chamber through the load lock station. We used a rotating wheel sample holder to prepare a set of samples with different layer sequences in a single vacuum cycle. Next, if necessary, the deposition of CoO$_x$ was performed in two steps: (I) Metallic Co was deposited on the substrate and (II) the substrate was moved into the load lock station and exposed to 100 mbar of O$_2$ gas for 2 h. For the samples with a rough S layer, the remaining sequence of layers (F1/N/F2/AD/S) was deposited at a substrate temperature of $T_{sub}$ = 300 K. In order to prepare samples with a smooth S layer, we set the substrate temperature to $T_{sub}$ = 150 K. The compositions of the samples are summarized in Table 1. We used the following deposition rates: 0.5 Å/s for the Py and Cu layers; and 2 and 12 Å/s for the rough and smooth Pb films, respectively.

**Table 1** Composition of samples with rough (i) and smooth (ii) S layer

| Type | No. | Composition |
|---|---|---|
| (i) | 1 | Py(5)Pb($d_{Pb}$), $d_{Pb}$ = 60, 70, 75, 90, 105, 150, 240 nm |
| | 2 | Py(5)Cu(2)Pb($d_{Pb}$), $d_{Pb}$ = 70, 90, 200 nm |
| | 3 | CoO$_x$Py1(3)Cu(4)Py2(1)Cu(2)Pb(70) |
| (ii) | 1 | Py(5)Pb($d_{Pb}$), $d_{Pb}$ = 70, 95, 120 nm |
| | 2 | Py(5)Cu(2)Pb($d_{Pb}$), $d_{Pb}$ = 47, 70, 80, 94, 110, 125, 155 nm |
| | 3 | CoO$_x$Py1(3)Cu(4)Py2(1)Cu(2)Pb(70) |

To confirm the thicknesses of the layers, as well as the interface roughness and the morphology of the Pb layer, cross sections of the samples were investigated using a transmission electron microscope (TEM) (FEI, Tecnai F30) operating at an acceleration voltage of 300 kV. The electron-transparent lamellas were prepared through the focused ion beam technique using a Zeiss 1540XB cross beam machine. After inspection of the imaging, a C/Pt–O–H protection layer (technical layers in Fig. 2) was deposited at the position of interest. The lamella was cut using a focused 30 keV Ga ion beam and after being lifted out, it was welded onto an electron microscopic girder. The protection layer reduced Ga implantation in the sample region close to the surface. The cross sections were analyzed through conventional fixed-beam imaging in the TEM, using its high-resolution option, as well as in a scanning TEM (STEM) mode using a high-angle annular dark field detector (HAADF). Additionally, using energy-dispersive X-ray spectroscopy in a synthesis mode, the existence of the thin Py layer could be confirmed. The interfaces between the single layers could be seen clearly in the TEM micrographs as well as in the STEM-HAADF images (Fig. 2). As can be seen in Figs. 2(b) and 2(c), in the MgO/Py/Pb structure prepared at $T_{sub}$ = 300 K, the Pb layer grew in the form of overlapping islands with an island size of 0.2–1 μm. In the case of the MgO/Py/Cu/Pb structure prepared at $T_{sub}$ = 150 K, the TEM image of the cross section reveals a smooth surface on the Pb layer (Fig. 2(e)). The thickness of the Pb layer can be estimated as $d_{Pb}$ = 70 nm. The low image contrast between the Cu and Py layers did not allow us to estimate the thicknesses of the Cu and Py layers separately. However, we found that the thickness of the whole Py/Cu part amounts to $d_{PyCu}$ = 8.6 ± 0.1 nm, which is in good agreement with the targeted thicknesses of Py and Cu, $d_{Py}$ = 5 nm and $d_{Cu}$ = 2 nm.

The superconducting properties of the samples were studied using a four-contact resistivity measurement method, using the B2902A precision source/measure unit from Keysight Technologies. The samples were mounted in a $^4$He cryostat from Oxford Instruments inserted between the poles of a high-homogeneity dipole electromagnet from Bruker. The magnetic field was measured with an accuracy of ±0.3 Oe using a Hall probe. The temperature of the sample was controlled using a 230-Ω Allen-Bradley resistor thermometer, which is particularly sensitive in the temperature range of

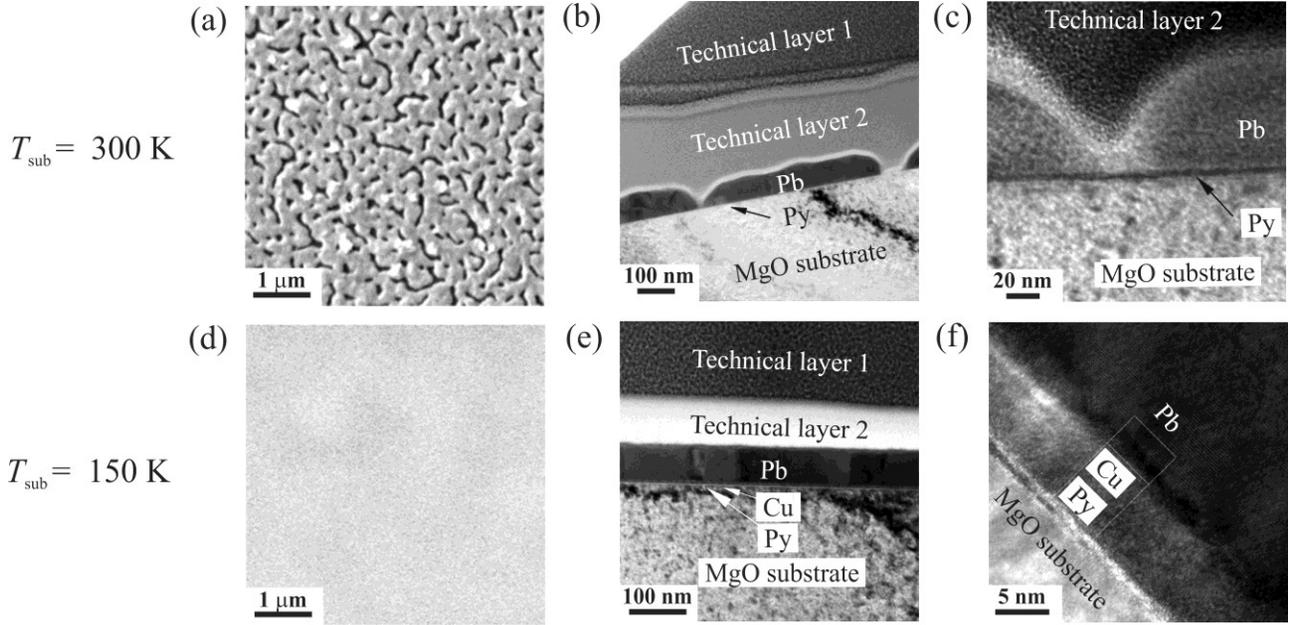

**Figure 2** Microscopic characterization of the samples with (top) rough and (bottom) smooth S layers deposited at substrate temperatures of 300 and 150 K, respectively. Micrographs of the surfaces of the Pb layers and TEM images of the cross sections obtained with the HAADF detector at two magnifications are shown, respectively, in panels (a), (b), and (c) for the Py(5)/Pb(70) structure with the rough Pb layer, and in panels (d), (e), and (f) for the Py(5)/Cu(2)/Pb(70) structure with the smooth Pb layer

interest.

We found that the residual resistivity ratio, $RRR = \rho(300\ K)/\rho(10\ K)$, of the studied samples lies in the interval $10 < RRR < 17$, with no notable difference with regard to the type of S layer. Figure 3 shows the characteristic superconducting transitions for the rough and smooth Py/Pb and Py/Cu/Pb structures. A comparison of Figs. 3(a) and 3(b) reveals that, for samples prepared at $T_{sub}$ = 300 K, the Cu AD layer does not influence the superconducting properties. The transition is sharp in both cases, with the same value for $T_c$. In contrast, the transition curve of the Py/Pb sample prepared at $T_{sub}$ = 150 K exhibits several steps, indicating degraded superconducting properties (Fig. 3(c)). Insertion of the Cu AD layer dramatically improves the quality of the transition (Fig. 3(d)), making it superior to those in Figs. 3(a) and 3(b). This result is similar to our earlier findings [15, 16].

All structures were magnetically characterized using a 7-T VSM SQUID magnetometer from Quantum Design. First, the samples were cooled from 300 to 10 K in the presence of an in-plane magnetic field of +4 kOe. At 10 K, the magnetic field was varied from +4 to −4 kOe and then back to +4 kOe. During this variation, the in-plane magnetic moment of the sample was measured (Fig. 5(a) and inset therein). In a strong positive field, the magnetizations of the F1 and F2 layers, $M_1$ and $M_2$, are aligned parallel and the sample is fully magnetized. Here, $M_2$ follows the sign change of the applied field and flips, giving rise to a step in the $M(H)$ curve (inset in Fig. 5(a)). Since $M_1$ is pinned to the AF layer, it stays against the field, and the AP configuration of the spin valve is realized. Eventually, $M_1$ is reversed in a field of −2.5 kOe, and the structure is fully magnetized in the opposite direction. It appears that, for most of the CoO$_x$/Py1/Cu/Py2/Cu/Pb samples, the saturation field for the Py2 film is of the order of 200 Oe (Fig. 5(a)). This means that choosing a switching field of $H_0$ = ±1,000 Oe to turn the sample from P to AP configurations and back is sufficient to sustain a homogenous magnetization for the Py2 layer that follows the switching field direction without the formation of a domain structure [17]. A very similar result was previously obtained for the CoO$_x$/Fe1/Cu/Fe2/Cu/Pb structure (Fig. 4 in Ref. [16]).

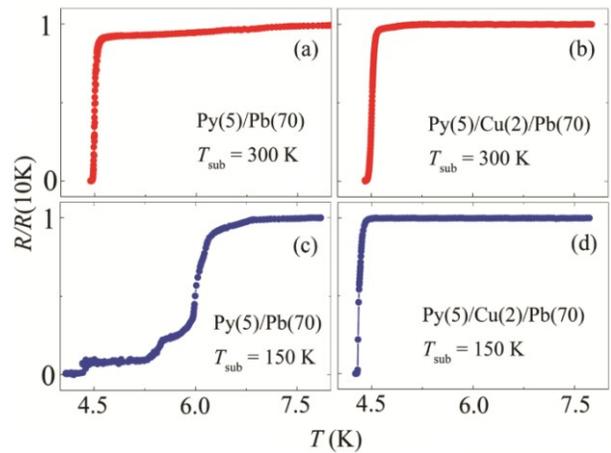

**Figure 3** Electrical transport characterization of the samples. Superconducting transition curves for samples Py(5)/Pb(70) and Py(5)/Cu(2)/Pb(70) with the rough Pb layer ($T_{sub}$ = 300 K) are shown in panels (a) and (b) and curves for the respective samples with the smooth Pb layers ($T_{sub}$ = 150 K) are shown in panels (c) and (d).

The influence of the ferromagnetic order on the superconducting properties of the S/F-based structures strongly depends on several parameters of the S layer. One of the most important is the thickness of the S layer, $d_S$. The thinner the S layer is, the more the superconductivity is suppressed, and, hence, the larger the magnitude of the SSVE that can be observed. The optimal thickness is typically deduced from the $T_c(d_S)$ dependence. Such a dependence can in principle be calculated theoretically (see, e.g., Ref. [18]). However, this theory does not consider the morphology of the S film. To address this issue, we measured the dependence of $T_c$ on the thickness of the Pb layer, $d_{Pb}$, for Py/Cu/Pb and Py/Pb heterostructures with rough and smooth Pb layers. The results are summarized in Fig. 4. Interestingly, the measured samples exhibit very similar $T_c(d_{Pb})$ dependence, suggesting that the Cu AD layer does not affect the S/F proximity effect and that the morphology of the S structure does not influence the suppression of $T_c$. We fit the experimental data with the theoretical model of Fominov et al. [18] using a reasonable set of parameters for that theory: coherence lengths for the S and F layers of $\xi_S$ = 41 nm and $\xi_F$ = 13 nm; an exchange field of $h$ = 0.3 eV; electron mean free paths for the S and F layers of $l_S$ = 30 nm and $l_F$ = 3 nm; and S/F boundary transparency parameters of $\gamma$ = 0.74 and $\gamma_b$ = 2.2. The fit agrees well with the experimental data, yielding a critical Pb thickness of $d_{Pb}^{cr}$ = 40 nm at which the superconductivity is completely suppressed (Fig. 4).

The results in Fig. 4 are helpful in determining the optimal thickness of the S layer for superconducting spin valve structures.
We chose $d_{Pb}$ = 70 nm as this is large enough to provide a measurable $T_c$, and yet is close to the critical thickness $d_{Pb}^{cr}$, which is favorable for the observation of the SSVE. We prepared superconducting spin valve samples of CoO$_x$(3)/Py1(3)/Cu(4)/Py2(1)/Cu(2)/Pb(70) with rough and smooth Pb layers. SQUID characterization (see Fig. 5(a) and the inset therein) did not reveal any difference in magnetic properties between these two systems. However, for the heterostructures with the rough S layer, we found no shift of $T_c$ when switching between the AP and P states, i.e., $\Delta T_c$ = 0, suggesting the absence of the SSVE (Fig. 5(b)). In contrast, for the spin valve system with the smooth S layer, the $\Delta T_c$ amounts to 100 mK (Fig. 5(c)).

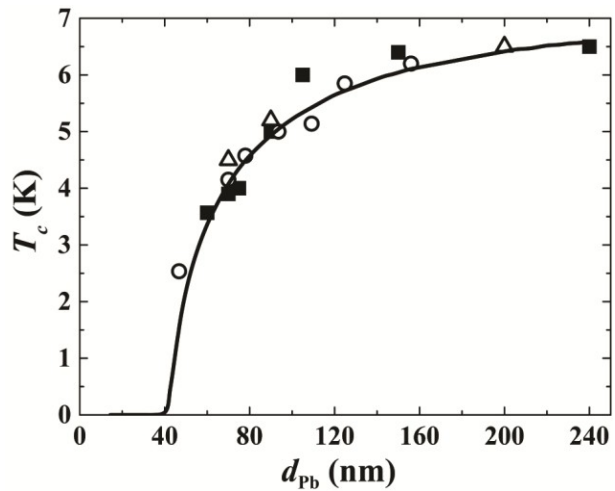

**Figure 4** Dependence of $T_c$ on the thickness of the superconducting Pb layer, $d_{Pb}$, for the Py/Pb and Py/Cu/Pb structures with a rough Pb layer (squares and open triangles, respectively) and for Py/Cu/Pb structures with a smooth Pb layer (open circles). The solid line denotes the theoretical fit.

The results in Figs. 5(b) and 5(c) clearly demonstrate the significant influence of the morphology of the S layer on the SSVE. Obviously, this is not caused by differences in the suppression of $T_c$ by the S layer, since the $T_c(d_S)$ dependence appears to be insensitive to the morphology (Fig. 4). In addition, the possibly smaller contac area area at the S/F interface in the case of the rough S layer, as compared to the smooth case, is not expected to play a role, since, according to theoretical models [5, 6], the size of the contact area does not affect the physical processes relevant to the SSVE. More likely, the structure of the S layer influences the oscillatory behavior of the electron pair wave function, which is inherent to the S/F proximity effect [19]. The superconducting condensate in the trilayer S/F1/F2 is described by Green's function in the S and F layers.

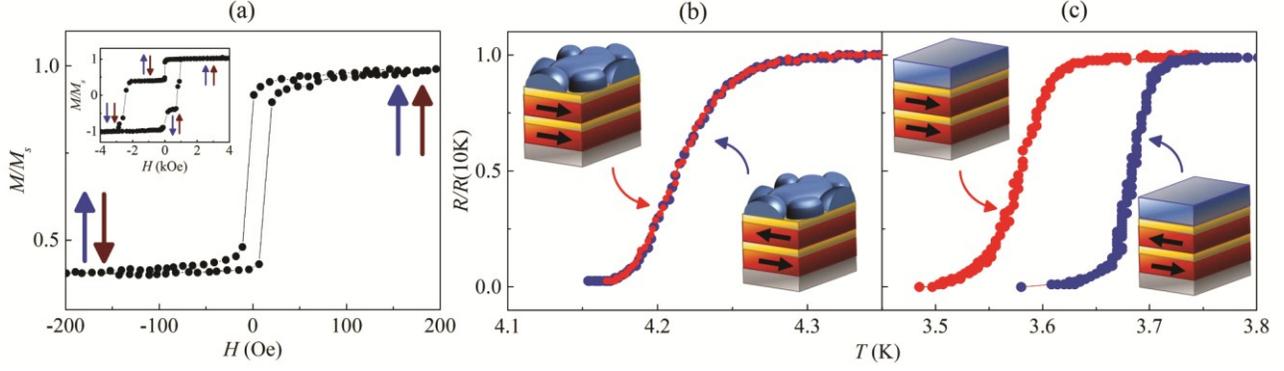

**Figure 5** Magnetic and superconducting properties of the spin valve samples with rough and smooth S layers. (a) A minor magnetic hysteresis loop corresponding to the magnetization reversal of the free $Py^2$ layer, and the major hysteresis loop (inset in panel (a)) for the sample $CoO_x$/Py1(3)/Cu(4)/Py2(1)/Cu(2)/Pb(70) with a smooth Pb layer. This magnetic behavior is also typical for the spin valve sample with a rough Pb layer. Superconducting transitions, measured for AP and P states for $CoO_x$/Py1(3)/Cu(4)/Py2(1)/Cu(2)/Pb(70) with (b) rough and (c) smooth Pb layers. Arrows depict the mutual orientations of the magnetizations of the Py1 and Py2 layers.

The Cooper pair penetrated into the F layer acquires a finite momentum, which cause spatial oscillation of the Green's function inside the F layer. The wave function reflected from the F surface opposite to the S/F interface can interfere with the incoming wave function. The SSVE is essentially related to the interference pattern that arises in this process [17, 20].

The in-plane inhomogeneity of the thickness of the S layer composed of overlapping islands of sizes 0.2–1 μm (Fig. 2) is considerable at distances much larger than the coherence length $\xi_S$ and the electron mean free path $l_S$ in the S layer, both on the order of some tens of nanometers. Naturally, such structural inhomogeneity can produce in-plane inhomogeneity of the Green's function in the S layer. Because the Green's functions in the S and F layers are connected through the S/F boundary conditions, the transferred electronic inhomogeneity can dampen interference in the F layer and thus suppress the SSVE.

In summary, the results of our electron microscopy study, in combination with transport and magnetic measurements, clearly demonstrated the crucial role of the morphology of the superconducting layer in the SSVE in metallic S/F1/F2 heterostructures. We argued that the in-plane inhomogeneity of the S layers "converts" into the inhomogeneity of the superconducting Green's function in the F layer, causing the suppression of the SSVE, whereas the dependence of $T_c$ on the thickness of the S layer remains unaffected by its surface morphology. The magnitude of the SSVE, $\Delta T_c$, can be tuned down to zero by increasing the roughness of the S surface and boosted up to $\Delta T_c$ = 100 mK by smoothening the S layer. This finding provides new insights into the sensitivity of the microscopic mechanism behind the SSVE to the real morphology of superconducting spin valves and can be important for the implementation of the SSVE in superconducting spin electronic devices.

## Acknowledgements

We gratefully acknowledge Ya. V. Fominov and M. Yu. Kupriyanov for fruitful discussions. This work was financially supported by the Deutsche Forschungsgemeinschaft (No. LE 3270/1-1). It was also partially supported by Russian Foundation for Basic Research (No. 14-02-00350-a) and the Program of the Russian Academy of Sciences.